\documentclass[useAMS,usenatbib,fleqn,twocolumn,published,letter]{aastex7}
\makeatletter
\let\frontmatter@title@above=\relax
\makeatother
\usepackage{multirow}
\usepackage{graphicx}
\usepackage{xcolor,soul}
\usepackage{physics}
\usepackage{amssymb,amsmath}
\usepackage{xspace}
\usepackage{tabularx}
\usepackage[T1]{fontenc}

\def\sgra{Sgr~A$^*$\xspace}
\def\m87{M87$^*$\xspace}

\begin{document}

\shorttitle{Millimeter to X-ray in AGN}
\title{On the Physical Origins of the Millimeter Fundamental Plane in Active Galactic Nuclei}

\shortauthors{Mazde et al.}
\correspondingauthor{Kratika Mazde}
\email{kratika.mazde@iap.fr}

\author[0000-0002-6232-4743]{Kratika Mazde}
\affiliation{Institut d'Astrophysique de Paris, CNRS \& Sorbonne Universit\'e, UMR 7095 98 bis boulevard Arago, 75014, Paris, France}
\affiliation{Center for Astrophysics | Harvard \& Smithsonian, 60 Garden Street, Cambridge, MA 02138, USA}
\affiliation{Black Hole Initiative, Harvard University, 20 Garden Street, Cambridge, MA 02138, USA}
\email{kratika.mazde@iap.fr}

\author[0000-0001-5287-0452]{Angelo Ricarte}
\affiliation{Center for Astrophysics | Harvard \& Smithsonian, 60 Garden Street, Cambridge, MA 02138, USA}
\affiliation{Black Hole Initiative, Harvard University, 20 Garden Street, Cambridge, MA 02138, USA}
\email{angelo.ricarte@cfa.harvard.edu}

\author[0000-0001-6952-2147]{George N.~Wong}
\affiliation{Princeton Gravity Initiative, Princeton University, Princeton, NJ 08544, USA}
\affiliation{School of Natural Sciences, Institute for Advanced Study, 1 Einstein Drive, Princeton, NJ 08540, USA}
\email{gnwong@ias.edu}

\begin{abstract}
Observations of active galactic nuclei have revealed a correlation between millimeter luminosity, X-ray luminosity, and mass, suggesting the emission in each of these bands is powered by a common source. Starting with a set of five general relativistic magnetohydrodynamic simulations with dynamically important magnetic fields, we perform ray-tracing calculations to produce spectra including synchrotron emission, bremsstrahlung emission, and Compton scattering.  Our models with similar Eddington ratios to the objects for which the relationship was inferred naturally reproduce observations without tuning. Our lower Eddington ratio models depart from this relationship, likely attributable to an observational bias against extremely low accretion rates. We find that inverse Compton scattering dominates the production of X-rays over bremsstrahlung radiation in almost all models, and in all models consistent with the observed correlation. We find only a modest spin dependence in this relationship. This study demonstrates that a compact, hot accretion flow with dynamically important magnetic fields can naturally explain observed millimeter and X-ray properties in low-luminosity active galactic nuclei. Future work should explore the impacts of non-thermal electron populations, weaker magnetic fields, and radiative cooling.
\end{abstract}


\section{Introduction}
\label{sec:introduction}

Supermassive black holes are found at the centers of most massive galaxies, where they accrete gas and impart energy to their hosts \citep{Kormendy&Ho2013, Heckman&Best2014}.  When accreting, these Active Galactic Nuclei (hereafter AGNs) can outshine their host galaxies across most of the electromagnetic spectrum \citep[e.g.,][]{Netzer:2015jna,Hickox&Alexander2018}. The emission mechanisms responsible for the different parts of the spectral energy distribution (SED) spectra are imperfectly understood, and thus developing physically motivated models of AGN emission is essential for constraining the underlying accretion physics.

One interesting observational constraint is the classic ``fundamental plane'' (hereafter FP), an empirical correlation between the supermassive black hole mass ($M_{BH}$), 5 GHz radio luminosity ($L_{5, \rm GHz}$) and 2-10 keV X-ray luminosity
($L_{X, 2-10 keV}$) that extends over many orders of magnitude  \citep{Corbel+2003,Gallo+2003,Merloni+2003,Kording+2006,Plotkin+2012,Wang+2024,Long+2025}. This has been proposed as evidence for scale-invariant coupling between radiatively inefficient accretion disks and compact jets, where X-ray emission originates from the inner hot disk and radio emission from compact jets, modulated by black hole mass \citep{Markoff+2003,Falcke+2004,Fender+2004,Gultekin+2019}. 

More recently, similar correlations have been found between millimeter and X-ray wavelengths, mostly for high-luminosity samples\citep{Behar+2015,Behar+2018,Kawamuro+2022,Ricci+2023,Venselaar+2026}.  Temporal correlations observed between the two bands suggest a common and compact emission region \citep{Baldi+2015,Behar+2020,Petrucci+2023,Shablovinskaya+2024,Jana+2025}. By including mass, and extending to lower specific accretion rates, \citet{Ruffa+2024} identified a ``Millimeter Fundamental Plane'' (hereafter mmFP), a relationship between black hole masses $M_{BH}$, nuclear ($\ll 100\,{\rm pc}$) luminosities $L_{\nu, mm}$ and intrinsic X-ray luminosities $L_{X, 2-10 keV}$.   

In the millimeter, general relativistic magnetohydrodynamics (GRMHD) simulations have been remarkably successful at matching resolved observations of accretion flows by the Event Horizon Telescope (EHT), even when including only thermal electron distribution functions \citep{EHTC+2019e,EHTC+2021b,EHTC+2022e,EHTC+2023,EHTC+2024c}.  Comparable models for radio jets require large simulation domains, often ad-hoc prescriptions for non-thermal electron distributions, and may be affected more strongly by environmental interactions \citep{Moscibrodzka+2016,Romero+2017,Blandford+2019}.  Consequently, the mmFP may be less affected by uncertain jet and plasma astrophysics and may probe smaller scales compared to the traditional fundamental plane using 5 GHz, making them more accessible to GRMHD modeling.

In modeling emission in these bands, the three most important radiative processes to consider are synchrotron emission (dominant in the radio/millimeter), and both bremsstrahlung and inverse-Compton scattering (which contribute to the X-rays). \citet{delPalacio+2025} develop a one-zone corona emission model with a hybrid thermal and non-thermal electron population \citep{Inoue+2019} and fit the spectral energy distributions (SEDs) of 7 radio-quiet AGNs.  Their modeling infers for the corona modest non-thermal energy density fractions (0.5-10\%), magnetic field strengths of 10-60 G, and radii of 60-250 gravitational radii.  Meanwhile, \citet{Hankla+2025} propose that millimeter emission in radio-quiet AGNs arises from an extended outflow around $\sim$$10^{4}$ gravitational radii, preferring larger magnetic field strengths of $\sim$$10^{3-4} \ \mathrm{G}$.

In this work, we investigate the mmFP by calculating SEDs using GRMHD simulations applicable to low-luminosity AGN for the first time.   In \autoref{sec:methodology}, we introduce the accretion flow simulations used in this study and the spectral modeling performed in this work.  In \autoref{sec:results} we present the results of this study, spectral energy distributions from the millimeter to the X-ray across a grid of accretion flow parameters.  We discuss the implications and limitations of our study in \autoref{sec:discussion}, and summarize our conclusions in \autoref{sec:conclusion}.

\section{Methodology}
\label{sec:methodology}

Starting with 5 “Magnetically Arrested Disk” (MAD) GRMHD simulations, we calculate SEDs as a function of SMBH mass $M_\bullet$, accretion rate $\dot m$, spin $a_\bullet$, and asymptotic ion-to-electron temperature $R_\mathrm{high}$.

\subsection{GRMHD Simulations}

We use as our starting point the 3D GRMHD simulations described in \citet{Dhruv+2025}, which feature 5 spins, $a_\bullet \in \{-0.94, -0.50, 0.00, +0.50, +0.94 \}$.  These simulations have been well-studied in EHT theory analysis \citep{EHTC+2022e,EHTC+2024c}.  To limit the scope of our analysis, we only include ``Magnetically Arrested Disk'' (MAD) simulations, which are characterized by dynamically important magnetic fields that alter the structure of the inflow and can power efficient jets \citep{Bisnovatyi-Kogan&Ruzmaikin1974,Igumenshchev+2003,Narayan+2003,Tchekhovskoy+2011}.  When breaking degeneracies with either polarization or multi-wavelength information, EHT studies of \m87 and \sgra \citep{EHTC+2021b,EHTC+2022e,EHTC+2023,EHTC+2024c} favor MAD models over less strongly magnetized ``Standard and Normal Evolution'' (SANE) models.  Testing MAD models against the mmFP, derived from a more representative sample than the two EHT targets, is major motivation of this study.

The simulations were run using the {\sc KHARMA} code \citep{Prather2024} and were initialized with a \citet{Fishbone&Moncrief1976} torus of plasma with a pressure maximum at 41 $r_g$\footnote{We define $r_g=GM_\bullet/c^2$, where $G$ is the gravitational constant, $M_\bullet$ is the black hole mass, and $c$ is the speed of light.  We further define $t_g = r_g/c$.}, threaded with an initially weak poloidal magnetic field.  The resolution of these simulations is 288x128x128 in the radial, latitudinal, and azimuthal directions, and they are evolved to 30,000 $t_g$ assuming an adiabatic index of $4/3$.  In this work, we select 21 snapshots from each of these simulations evenly spaced in time between 15,000 $t_g$ and 30,000 $t_g$ to ensure a sampling over quasi-steady accretion. In the following analysis, these snapshots are each individually rescaled to different masses and accretion rates.

\subsection{SED Calculations}

Following the {\sc Patoka} pipeline \citep{Wong+2022}, we compute multi-wavelength spectra with the general relativistic Monte Carlo radiative transfer code \texttt{grmonty} \citep{Dolence+2009}, which is optimized for optically thin, hot plasma in strong gravitational fields. \texttt{grmonty} tracks photon emission, absorption, and scattering within the Kerr metric. The code includes synchrotron and bremsstrahlung emission as seed processes, and accounts for inverse Compton scattering in both the Thomson and Klein-Nishina regimes.

Emission is calculated in \texttt{grmonty} by discretizing the photon field into ``superphotons'' or ``photon packets'' generated with weight $w$, coordinates $x^{u}$, and wave vector $k^{u}$, in local orthonormal tetrads co-moving with the fluid.  
In a given time step $\Delta t$, superphotons over the whole volume of the model are produced according to
\begin{equation*}
    N_{s,tot} = \Delta t \int \sqrt{-g} \dd{x} \dd{\nu} \dd{\Omega} \frac{1}{w} \frac{j_\nu}{h\nu},
\end{equation*}
where $j_v$ is the fluid-frame emissivity, $\nu$ is the frequency, $\Omega$ is the solid angle in the fluid frame, and $h$ is the Planck's constant. To calculate $\nu$ and direction vectors ${\hat{n}}$, \texttt{grmonty} makes use of rejection sampling based on the local emissivity function. Emissivity modules are modular, allowing for different emission processes to be incorporated with minimal structural change.  We use radiative transfer coefficients appropriate for thermal electron distribution functions.

Simulations are post-processed under the ``fast light'' approximation, enabling rapid calculation of spectra for each snapshot.  
Photons are produced as synchrotron or bremsstrahlung radiation, and they may be scattered as they move along their geodesics. This results in the production of inverse Compton upscattered photons. As we shall explore, synchrotron emission dominates in the radio to sub-millimeter, while inverse Compton upscattering produces high-energy tails relevant for X-ray observational constraints.  The resulting SEDs serve as direct inputs for comparison with sub-millimeter and X-ray observations and allow for parameter estimation via forward modeling.

\begin{table}[htb]
\begin{tabular}{|l|l|}
\hline
\textbf{Parameter} & \textbf{Values} \\
\hline
$a_\bullet$              & -0.94, -0.5, 0, 0.5, 0.94                     \\
$M_\bullet$ $[M_\odot]$  & $10^6$, $10^7$, $10^8$, $10^9$, $10^{10}$ \\
$\dot{m}$   & $10^{-8}$, $10^{-7}$, $10^{-6}$, $10^{-5}$, \textcolor{gray}{$10^{-4}$}                            \\
$R_\mathrm{high}$ & 1, 40, 160                               \\
\hline
\end{tabular}
\caption{Summary of models calculated in this work.  All models are MAD and contain only thermal electron distribution functions.  Each parameter combination is simulated 21 times to sample variability.  Models with $\dot{m}=10^{-4}$ were found to almost always exhibit unphysically large radiative efficiencies and are therefore excluded from most of our analysis.}
\label{tab:parameters}
\end{table}

Our GRMHD simulations span 5 spin values, and we rescale them in post-processing to explore three more free parameters:  SMBH mass ($M_\bullet$), Eddington ratio ($\dot{m}$), and asymptotic ion-to-electron temperature ratios in weakly magnetized regions ($R_\mathrm{high}$).  Our parameter combinations are summarized in \autoref{tab:parameters}.  In our conventions, we define the Eddington ratio $\dot{m}$ via

\begin{equation}
    \dot{m} = \frac{\dot{M}_\bullet}{\dot{M}_\mathrm{Edd}},
\end{equation}

\noindent where 

\begin{equation}
    \dot{M}_\mathrm{Edd} = \frac{L_\mathrm{Edd}}{0.1c^2} = \frac{4\pi G M_\bullet m_p}{0.1 \sigma_T c},
\end{equation}

\noindent $m_p$ is the mass of the proton, and $\sigma_T$ is the electron cross-section to Thomson scattering.  In other words, $\dot{m}$ is the accretion rate relative to that of an Eddington-limited system converting mass to radiation with a fiducial 10\% efficiency.  True radiative efficiencies of a subset of our models obtained by integrating our spectra are presented and discussed in \autoref{sec:radiative_efficiency}.

The parameter $R_\mathrm{high}$ modulates the asymptotic ion-to-electron temperature ratio in each cell via

\begin{equation}
    \frac{T_i}{T_e} = R_\mathrm{high}\frac{\beta_p^2}{1+\beta_p^2} + \frac{1}{1+\beta_p^2} \label{eqn:Rhigh}
\end{equation}

\noindent where $T_i$ and $T_e$ are the ion and electron temperatures respectively, and $\beta_p$ is the ratio of gas to magnetic pressure.  This parameterization from \citet{Moscibrodzka+2016} qualitatively captures the behavior of simulations where electron heating and cooling is modeled self-consistently.  Heat is preferentially supplied to protons and electrons can more efficiently cool, leading to this temperature asymmetry, which is more pronounced in more weakly magnetized regions \citep{Rees+1982,Narayan&Yi1995, 1999ApJ...520..248Q,Howes2010,Kawazura+2019,Chael+2019,Dihingia+2023}.  EHT studies have preferred $R_\mathrm{high} \gg 1$ when accounting for both multi-wavelength and polarization constraints \citep{EHTC+2021b,EHTC+2022e,EHTC+2023,EHTC+2024c}.

These four parameters---spin, mass, accretion rate, and asymptotic ion-to-electron temperature---define a physically motivated 4-dimensional parameter space over which our SEDs are computed.  We sample 21 independent snapshots for each parameter combination to probe AGN variability and compute time averages.  A minority of snapshots are discarded due to numerical artifacts in the radiative post-processing, identified through visual inspection.

\section{Results}
\label{sec:results}

\subsection{Comparison with the mmFP}
\label{sec:models_vs_data}

\begin{figure}
\includegraphics[width=0.5\textwidth]{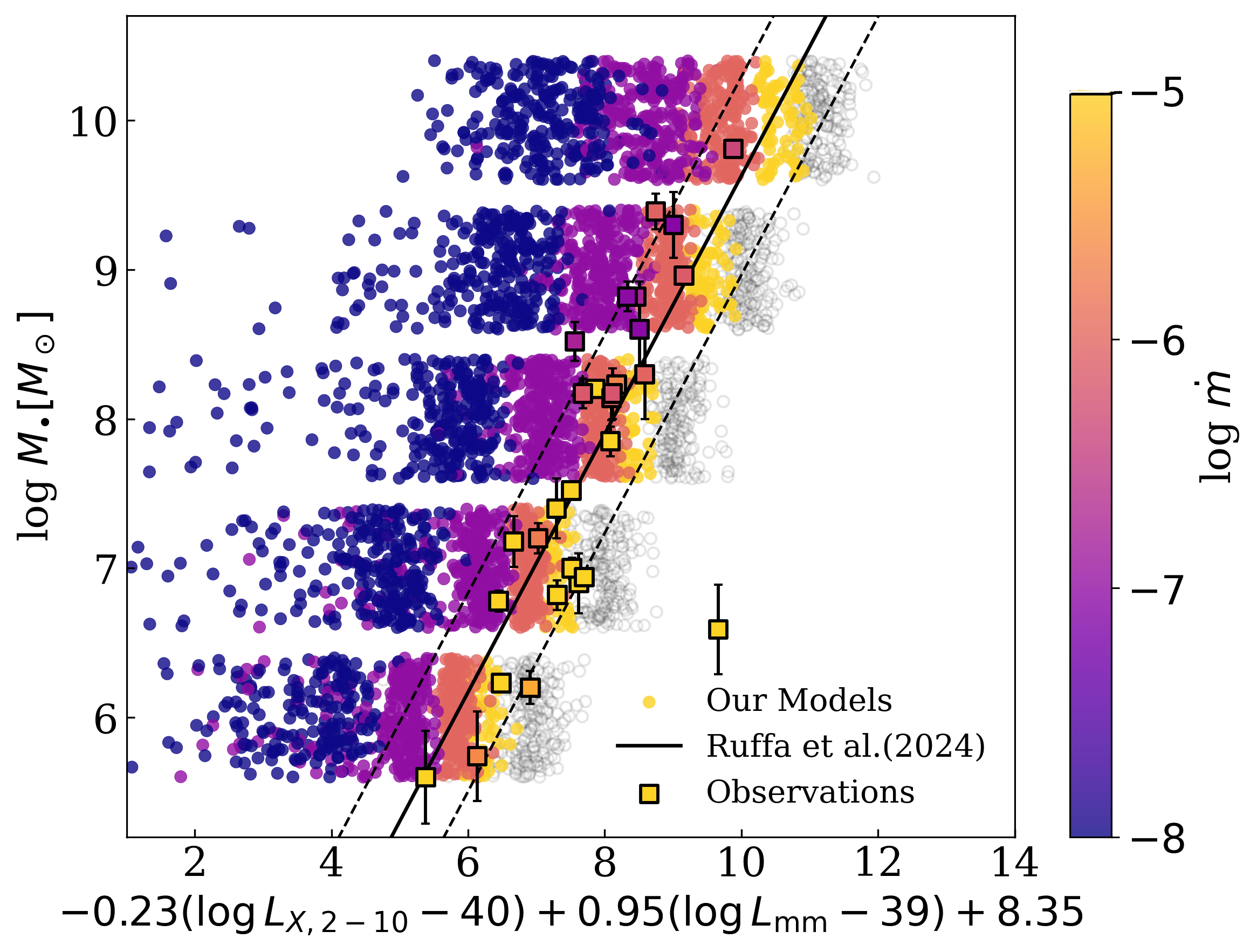}
\caption{Comparison of our models with the ``Millimeter Fundamental Plane'' reported in \citet{Ruffa+2024}.  Each point represents one GRMHD simulation, color-coded by $\dot{m}$.  The best fit from \citet{Ruffa+2024} is visualized with black lines, as well as the values from their primary sample of AGN.  We find that our GRMHD models with $\log \dot{m} \lesssim -7$ drift to the left of the relation. Models with $\dot{m}=10^{-4}$, shown here as gray circles, were almost all found to have $\epsilon_{\rm rad} \gg 0.1$; these are excluded from future analysis.}
\label{fig:mmFP}
\end{figure}

In \autoref{fig:mmFP}, we visualize our models on the millimeter fundamental plane proposed by \citet{Ruffa+2024}.  Note that for plotting, we have artificially added scatter to the black hole masses of our simulations (sampled discretely at the 5 values listed in \autoref{tab:parameters}) to improve readability.  Models are color-coded according to $\dot{m}$.  In gray, we show our $\dot{m}=10^{-4}$ models, which almost all exhibit $\epsilon \gg 0.1$ and are excluded from future figures.  $L_{X,2-10}$ is obtained by integrating the specific luminosity $L_\nu$ between 2 and 10 keV, while we define $L_{\mathrm{mm}} = \nu L_{\nu,\mathrm{mm}}$. The relation found in \citet{Ruffa+2024} is overplotted, including the primary sample of SMBHs used for this regression.\footnote{These include the WISDOM and literature sources tabulated in their Table A1, but not the extended BASS sample.}

We find that our models with $10^{-6} \leq \dot{m} \leq 10^{-5}$ are consistent with the relation, but models with lower $\dot{m}$ drift to the left increasingly quickly as $\dot{m}$ decreases. As we will explain in \autoref{sec:explain_millimeter}, we attribute this behavior in our models to an optical depth effect that only occurs when the peak frequency moves to lower frequencies than the millimeter. Our $\dot{m}=10^{-4}$ models overshoot the relation to the right, and appear to imply that AGN with larger Eddington ratios would become increasingly inconsistent with the mmFP.  However, we find that the $\dot{m}=10^{-4}$ models are radiatively efficient due to inverse Compton scattering.  We expect that models that self-consistently account for the resultant cooling would attain significantly lower the temperatures at these accretion rates and higher, and thus the position of these models on the plot is unreliable.

This is to be interpreted with some care, since the mmFP of \citet{Ruffa+2024} is predominantly calibrated using nearby, low-luminosity AGN, with quasar-like sources considered only as an extension. The physical interpretation of the relation is therefore tied primarily to radiatively inefficient accretion flows, rather than to the radiatively efficient thin-disk regime usually associated with high-Eddington-ratio quasars. We therefore regard the displacement of our $\dot{m}=10^{-4}$ models as a limitation of applying models with no radiative cooling at high accretion rates, rather than as a prediction of mmFP breakdown.

\subsection{Physical Drivers of the Fundamental Plane}
\label{sec:physical_drivers}

\begin{figure*}
\centering
\includegraphics[width=0.32\textwidth,height=0.33\textwidth,keepaspectratio]{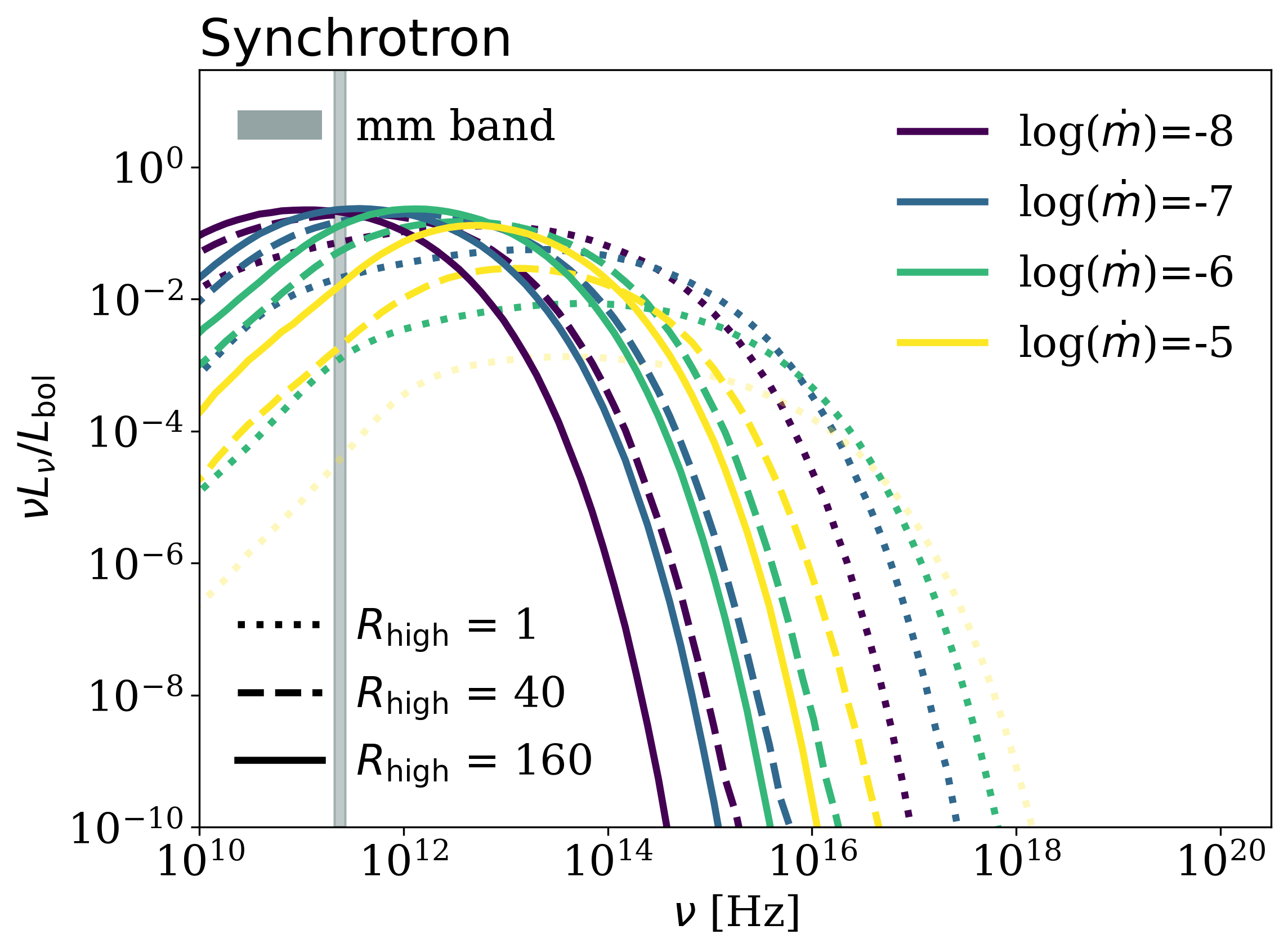}
\includegraphics[width=0.32\textwidth,height=0.33\textwidth,keepaspectratio]{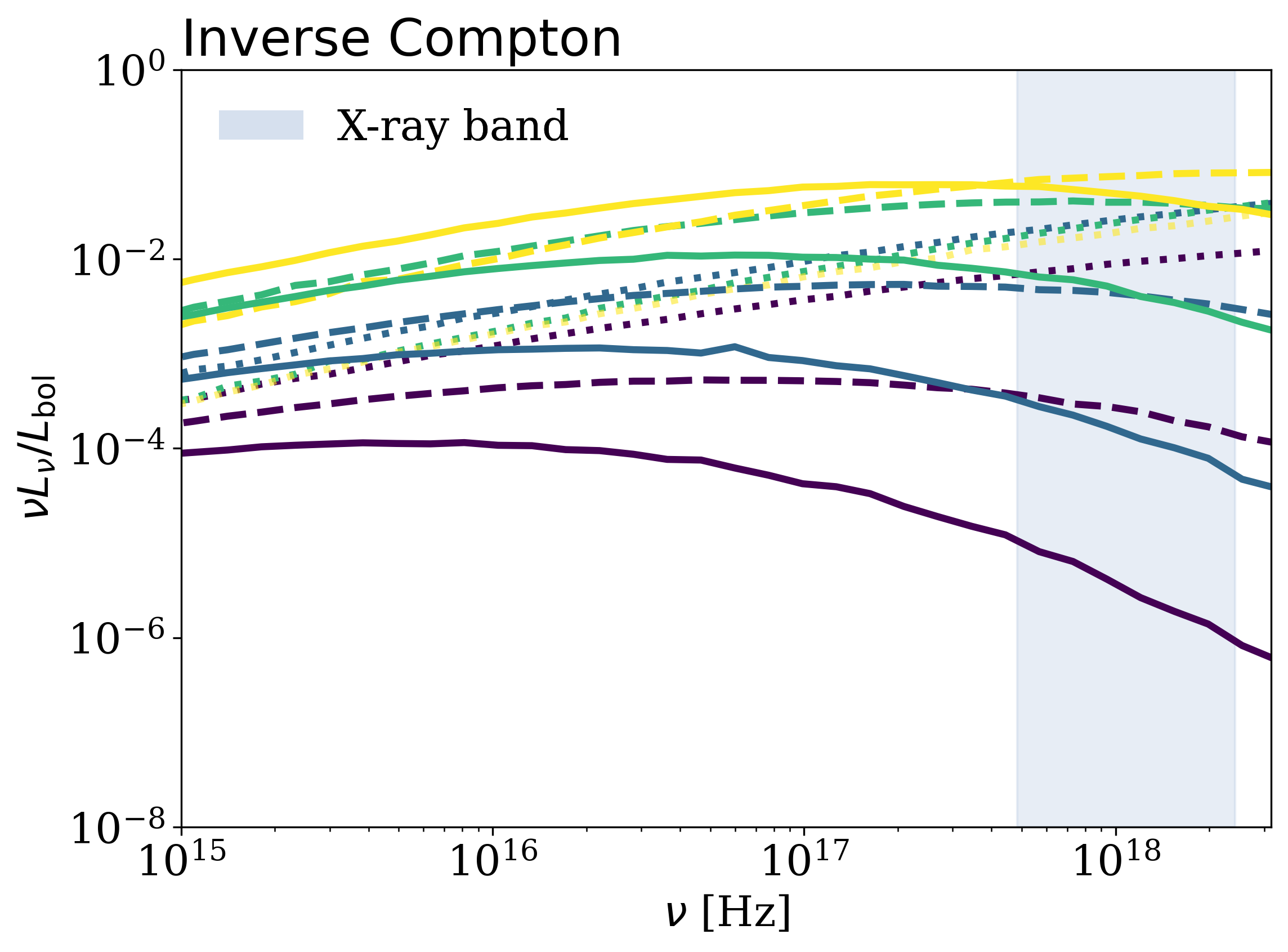}
\includegraphics[width=0.32\textwidth,height=0.33\textwidth,keepaspectratio]{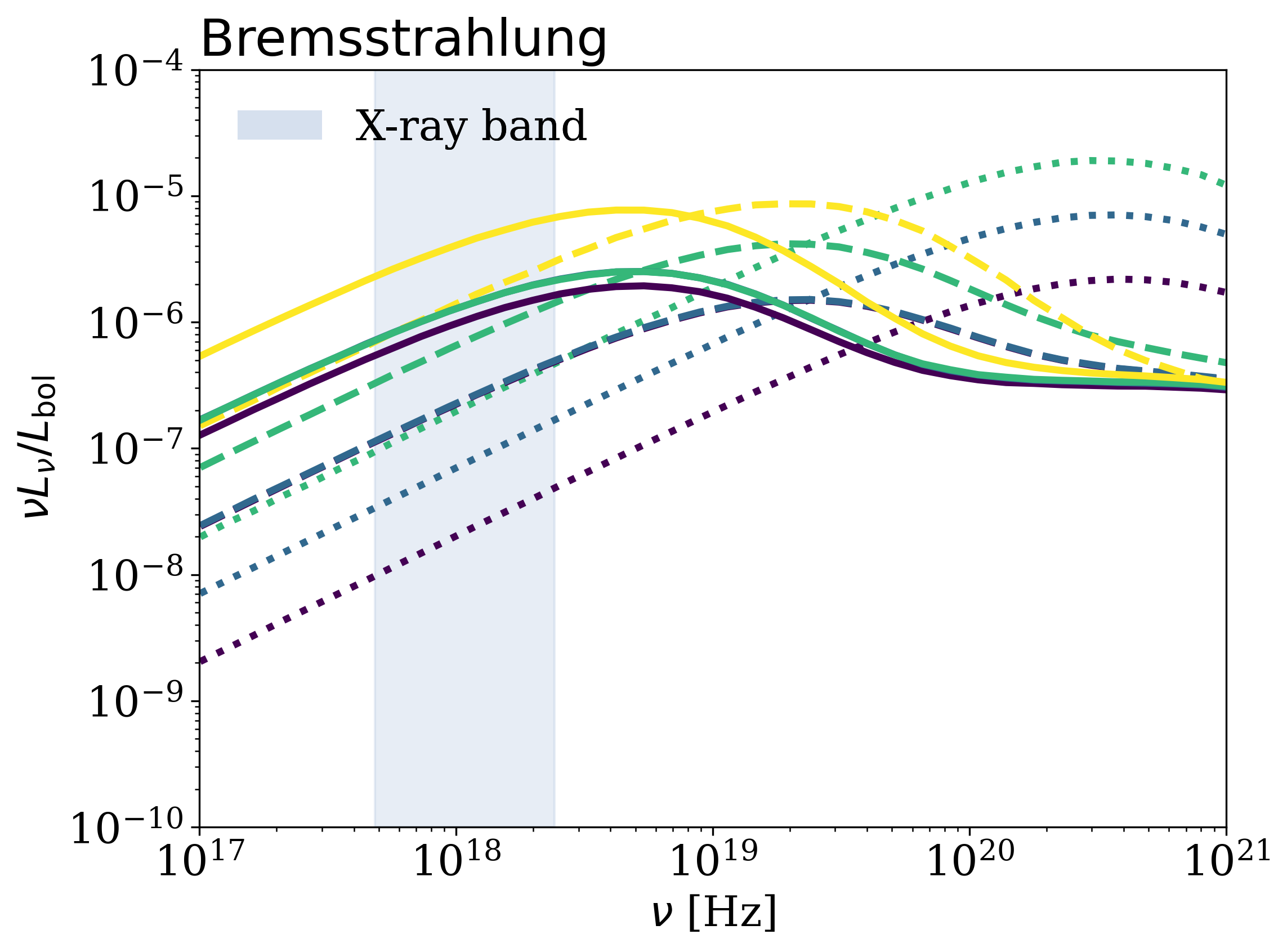}

\includegraphics[width=0.32\textwidth,height=0.33\textwidth,keepaspectratio]{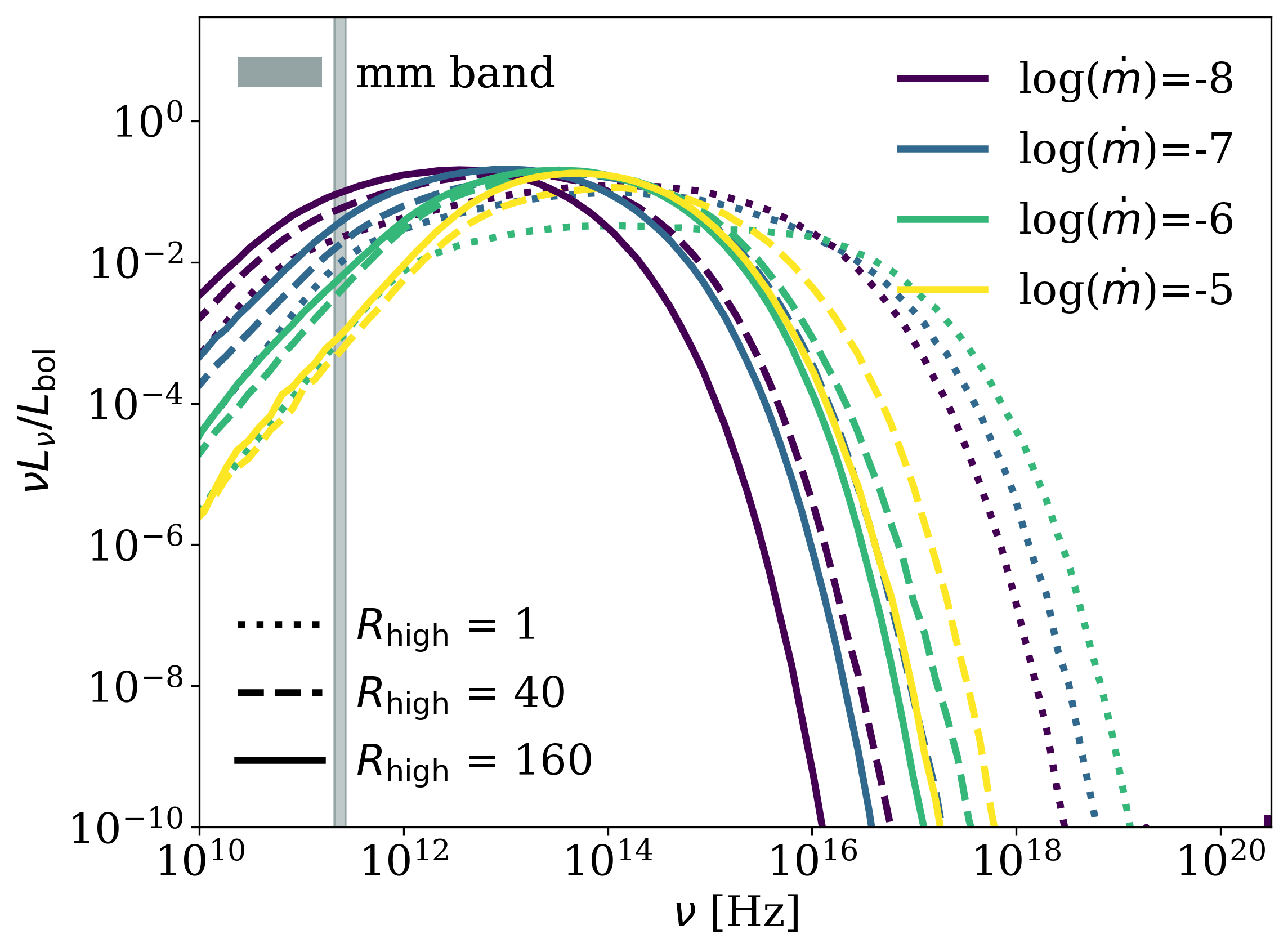}
\includegraphics[width=0.32\textwidth,height=0.33\textwidth,keepaspectratio]{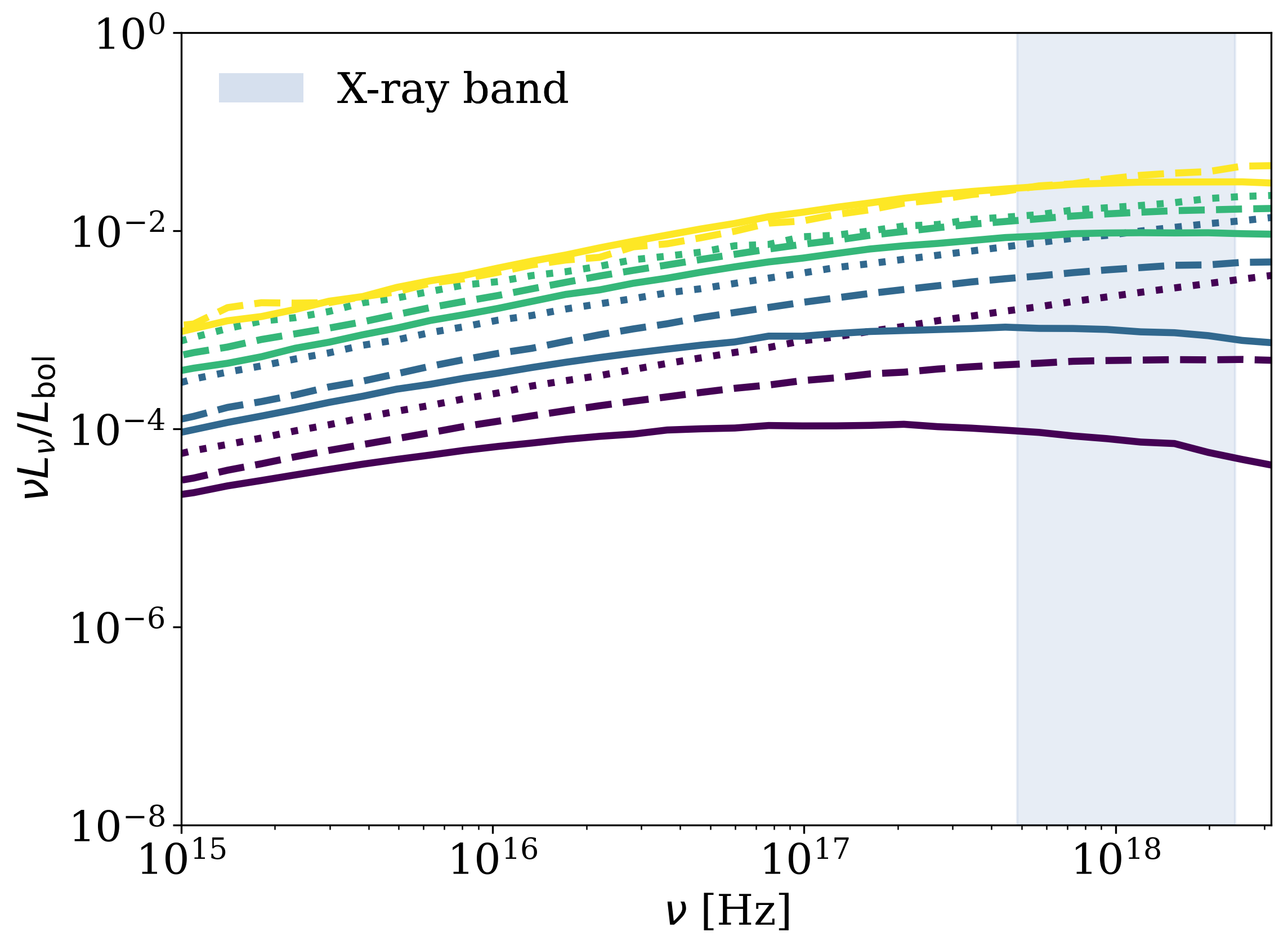}
\includegraphics[width=0.32\textwidth,height=0.33\textwidth,keepaspectratio]{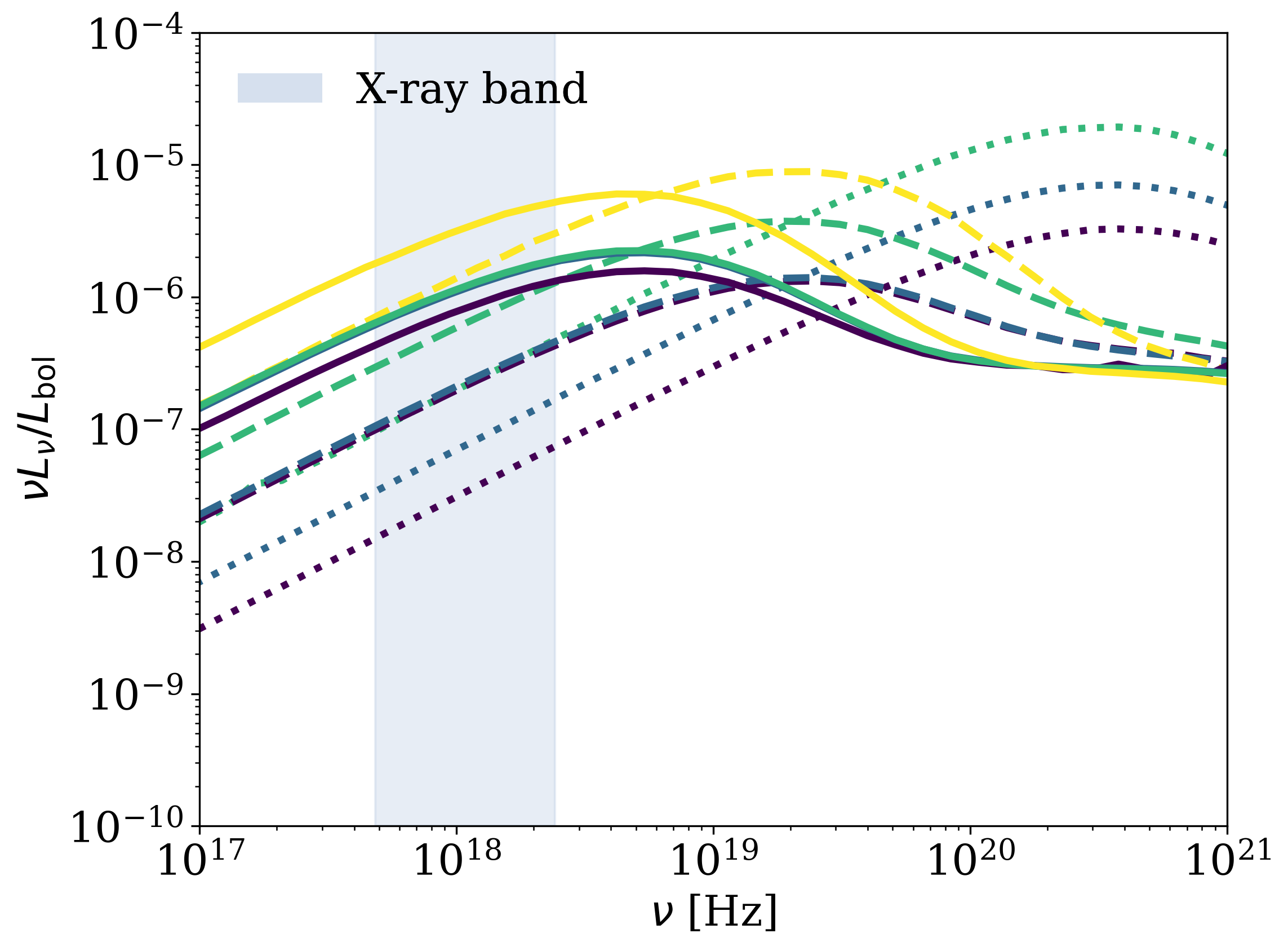}
\caption{Spectral energy distributions as a function of Eddington ratio $\dot{m}$ and $R_{\rm high}$ with fixed spin, $a_{\bullet}$=+0.94. Models with $M_{\rm BH} = 10^9 M_\odot$ are shown in the top row, and models with $M_{\rm BH} = 10^6 M_\odot$ are shown in bottom row.}
\label{fig:emission_breakdown}
\end{figure*}

In \autoref{fig:emission_breakdown}, we plot the evolution of synchrotron, Inverse Compton, and bremsstrahlung emission among our time-averaged SED models as a function of $M_\bullet$, $\dot{m}$, and $R_\mathrm{high}$.  For this and subsequent figures, we normalize our spectra by the bolometric luminosity, $L_\mathrm{bol}$, which is dominated by the synchrotron peak in our models.  ALMA band 6 is demarcated with a gray column, while the 2-10 keV X-ray band is demarcated with a blue column.  We fix $a_\bullet=0.94$ and show two masses, $10^9 \ M_\odot$ in the top row, and $10^6 \ M_\odot$ in the bottom row.

Here, we will break down the emission mechanisms and parameter dependencies in each band.  In interpreting these results, it is important to bear in mind the following scalings.  The density in GRMHD simulations scales as

\begin{equation}
    n \propto \dot{m}M_\bullet^{-1}, \label{eqn:density}
\end{equation}

\noindent the magnetic field strength scales as

\begin{equation}
    B \propto n^{1/2} \propto \dot{m}^{1/2}M_\bullet^{-1/2}, \label{eqn:magnetic_field}
\end{equation}

\noindent and $T_e$ evolves according to \autoref{eqn:Rhigh}, which behaves as

\begin{equation}
    T_e \propto\begin{cases}
			R_\mathrm{high}^{-1}, & \text{for $\beta \gg 1$}\\
            R_\mathrm{high}^{0}, & \text{for $\beta \ll 1$} 
		 \end{cases}
    \label{eqn:temperature}
\end{equation}

\noindent without a dependence on $\dot{m}$ or $M_\bullet$.  Any dependence on $a_\bullet$ is indirect, through emergent evolution in $n$, $B$, and $T_e$.

\subsubsection{Millimeter: Sourced by Synchrotron}
\label{sec:explain_millimeter}

Shown in the leftmost column of \autoref{fig:emission_breakdown}, synchrotron emission dominates the millimeter band for all of our models.  The turnover frequency increases as $\dot{m}$ increases, decreases as $R_\mathrm{high}$ increases, and increases as $M_\bullet$ increases.  All of these behaviors can be understood through the scaling of the critical frequency for synchrotron emission, which follows

\begin{align}
    \nu_\mathrm{synch} &\propto T_e^2B \propto T_e^2\dot{m}^{1/2}M_\bullet^{-1/2}
\end{align}

\noindent using \autoref{eqn:magnetic_field}.  At lower frequencies, these spectra follow the Rayleigh-Jeans law $L_\nu \propto \nu^2$ since we use a thermal eDF, while power-law eDFs would exhibit $L_\nu \propto \nu^{5/2}$.  At higher frequencies, these spectra inherit the exponential cutoff of the thermal eDF, which would fall off more gently if we had considered non-thermal electrons \citep{Ozel+2000, Ricarte+2023a}.  
In the optically thin limit, the total power emitted by synchrotron radiation follows $L \propto nB^2V$, where $V\propto M_\bullet^3$ is the volume of the emitting region.  Following our scalings, we can expect $L \propto \dot{m}^2M_\bullet$.  This strong scaling with $\dot{m}$ explains the clear $\dot{m}$ dependence in \autoref{fig:mmFP}, but may seem at odds with the \citet{Ruffa+2024} sample, where they report $L_{mm}$ mostly encodes $M_\bullet$.  This can be explained if for the objects in this sample, typically $\nu_\mathrm{synch} \gtrsim 230 \ \mathrm{GHz}$.  In this case, the emission transitions to optically thick at this frequency, and $L_\mathrm{mm}$ begins to scale with the radius $R$ of the 230 GHz photosphere.  For an optically thick ball of gas, $L_\mathrm{mm} \propto R^2 \propto M_\bullet^2$, but the actual scaling would depend in detail on the radial profiles of $n$, $B$, and $T_e$.  We discuss how we expect the mmFP to break down at low $\dot{m}$ in \autoref{sec:selection_effects}.

\subsubsection{X-ray: Dominated by Inverse Compton}
\label{sec:explain_Xray}

In the second and third columns of \autoref{fig:emission_breakdown} we plot the Inverse Compton ``IC'' and bremsstrahlung components of the spectrum respectively.  IC clearly dominates the X-ray emission in \textit{all} of our models, even for $M_\bullet=10^6 \ M_\odot$ and $\dot{m}=10^{-8}$.  For some of our $\dot{m}=10^{-5}$ models, the total IC luminosity becomes comparable to synchrotron.  All of our models are in the single-scattering regime, and the typical IC photon will be up-scattered to frequency $\nu_\mathrm{IC} \approx \frac{4}{3}\langle \gamma^2 \rangle \nu_\mathrm{seed}$, where $\gamma^2 \propto T_e^2$.  The seed photons come from synchrotron, and thus $\nu_\mathrm{IC}$ inherits the scaling of $\nu_\mathrm{synch} \propto T_e^2B$.  As a result,

\begin{equation}
    \nu_\mathrm{IC} \propto T_e^4 \dot{m}^{1/2}M_\bullet^{-1/2}.
    \label{eqn:nu_ic}
\end{equation}

\noindent This makes the X-ray part of the spectrum extremely sensitive to $R_\mathrm{high}$.  When these models were considered in \citet{EHTC+2022e}, an upper limit on the X-ray emission of \sgra led to a preference for larger values of $R_\mathrm{high}$, consistent with these findings.  The large $R_\mathrm{high}$ may itself be a consequence of electron cooling due to IC scattering.

The other intrinsic source of emission in our simulations is bremsstrahlung, or ``free-free'' emission.  Unlike synchrotron emission, bremsstrahlung is insensitive to the magnetic field.  The third column demonstrates that for almost all of our models, bremsstrahlung contributes negligibly to the bolometric luminosity of the system, although it is worth exploring physical trends.  Here, emission peaks at $\nu_\mathrm{bremss} \sim k_BT_e/h \sim 10^{20} \ \mathrm{Hz}$, in the gamma rays for $T_e \sim 1$.  As $R_\mathrm{high}$ increases, the single broad peak splits into two.  We suspect that the $R_\mathrm{high}$ prescription splits the accretion flow into a $T_e\sim 1$ jet funnel component and a $T_e\sim 1/R_\mathrm{high}$ disk component.

Bremsstrahlung has been shown to dominate the X-ray emission in other GRMHD models, particularly SANEs \citep{Yarza+2020,EHTC+2022e,EHTC+2024a}.  This is expected as SANE models often have larger values of $\dot{m}$ by orders of magnitude and lower values of $T_e$ by around an order of magnitude, and 
\begin{equation}
  j_{\nu, \mathrm{bremss}} \propto n^2 T_e^{-1/2} \propto \dot{m}^2M_\bullet^{-2}T_e^{-1/2}.
  \label{eq:brem_emissivity}
\end{equation}
The scaling of the IC luminosity with accretion rate can be understood in the context of hot accretion flow models. In the optically thin, single-scattering regime relevant for our models, the inverse Compton luminosity is given by
\begin{equation}
    L_{\rm IC}\simeq yL_{\rm syn},
\end{equation}
where $y=4\Theta_e\max(\tau,\tau^2)$ is the Compton parameter.  Since our models
are optically thin, $y\simeq4\Theta_e\tau$. At fixed dimensionless radius
$R/R_{\rm g}$, the density and size scale as
$n_e\propto \dot M/M_\bullet^2$ and $R\propto M_\bullet$, so that $\tau\propto n_e R \propto \frac{\dot M}{M_\bullet}\propto \dot m$. The synchrotron luminosity scales as
\begin{equation}
    L_{\rm syn}\sim n_e B^2R^3
    \propto
    \frac{\dot M^2}{M_\bullet}
    \propto
    \dot m^2 M_\bullet .
\end{equation}
Additionally, the synchrotron photon and magnetic energy densities follow
\begin{equation}
    \frac{U_{\rm ph}}{U_B}
    \sim
    \frac{L_{\rm syn}/R^2c}{B^2/8\pi}
    \propto
    \frac{\dot M}{M_\bullet}
    \propto
    \dot m ,
\end{equation}
while for constant magnetization, $U_B \sim B^2 / 8 \pi \propto n_e$. Taken together, these scalings yield $
L_{\rm IC}\propto yL_{\rm syn}\propto \dot m^3M_\bullet$ \citep[see also][]{Yuan+2003,Yuan&Narayan2014}. This steeper scaling with $\dot{m}$ compared to Bremsstrahlung implies that Bremsstrahlung dominates X-ray emission at the lowest $\dot{m}$ values.

\subsection{Only a Modest Dependence on Spin}
The classic fundamental plane, where radio emission plays the role of millimeter emission, is most strongly established for radio-loud AGN \citep{2004NewAR..48.1157F,Gultekin+2019}.  Radio emission originating from a jet may attain its energy through the \citet{Blandford&Znajek1977} mechanism, wherein the jet efficiency follows $\eta_\mathrm{jet} \propto a_\bullet^2 B^2$. 
A spread in the spins of the SMBH population is proposed as an explanation for variations in radio-loudness \citep{2007ApJ...667..704V}, as is a spread in magnetic field strengths \citep{Sikora&Begelman2013}.  

\begin{figure*}
    \centering
    \includegraphics[width=\textwidth]{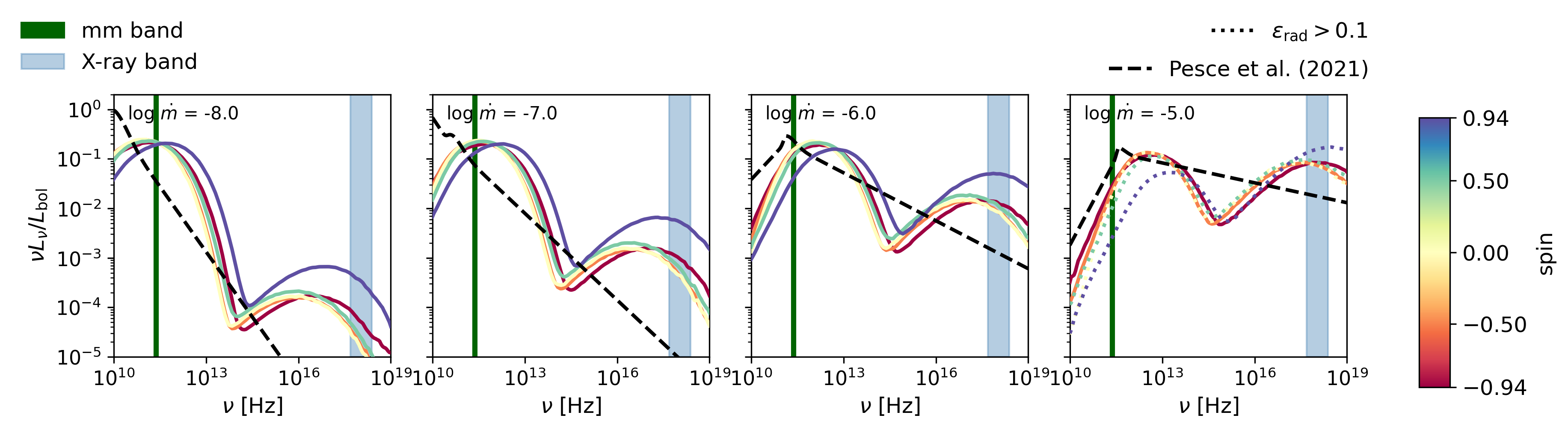}
    \includegraphics[width=\textwidth]{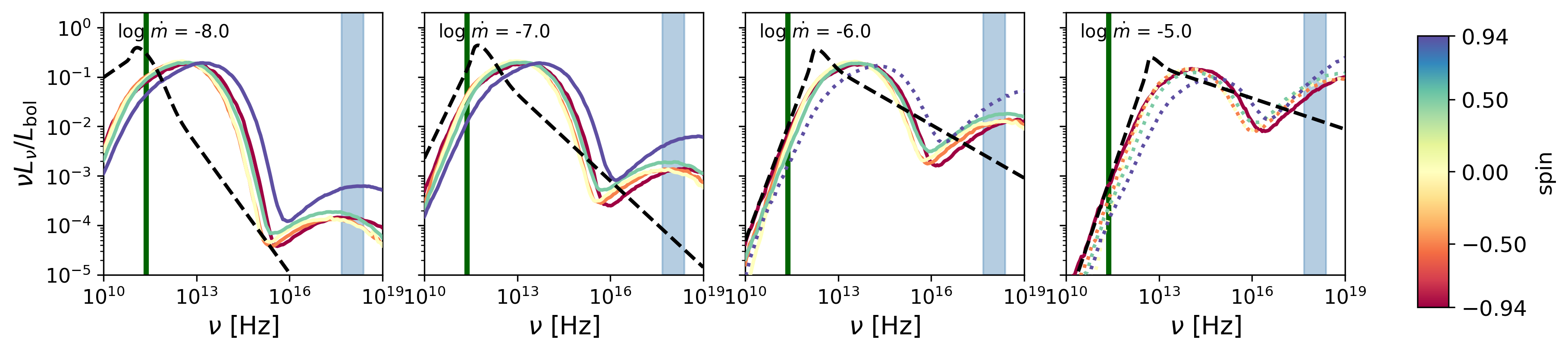}
    \caption{A grid of time-averaged normalized total spectra as a function of two parameters, Eddington ratio $\dot{m}$, spin $a_{\bullet}$, with fixed $R_{high}=40$.  Models with $M_{\rm BH} = 10^9 M_\odot$ are shown in the top row, and models with $M_{\rm BH} = 10^6 M_\odot$ are shown in bottom row respectively. The black dashed curve visualizes the semi-analytic ADAF/LLAGN model of \citet{Pesce2021}.}
    \label{fig:total}
\end{figure*}

Fixing $R_\mathrm{high}=40$, we plot our total SEDs as a function of $M_\bullet$, $\dot{m}$, and $a_\bullet$ in \autoref{fig:total}.  We would expect millimeter emission sourced by \citet{Blandford&Znajek1977} jet power to exhibit a similarly strong dependence on $a_\bullet$.  Instead, we find only a modest dependence on spin, generally consistent with EHT studies using these models \citep{EHTC+2022e}.  The most pronounced spin evolution occurs between spins 0.5 and 0.94. These $a_\bullet = 0.94$ models exhibit hotter jet funnels \citep[see Fig.~7 of][]{Dhruv+2025} than their lower spin counterparts, to which we attribute the higher synchrotron frequencies and IC normalizations. 

For comparison, we also show the semi-analytic ADAF/LLAGN model of \citet{Pesce+2021}. This model describes the total emission from radiatively inefficient accretion flows, including synchrotron and inverse-Compton emission, and is overlaid here as the black dashed curve using its default parameters. The ADAF/LLAGN model roughly matches some of the trends we see in our models: increasing the accretion rate increases the turnover frequency and the X-ray to radio luminosity ratio. However, the placement of these features can be offset by over an order of magnitude, and because the ADAF/LLAGN model does not produce a double-peaked SED structure, it also systematically underestimates the fraction of emission in the X-ray to a similar degree.

\subsection{Which models lie on the relation?}
\begin{figure*}
    \centering
    \begin{tabular}{cc}
    \includegraphics[width=0.9
    \textwidth]{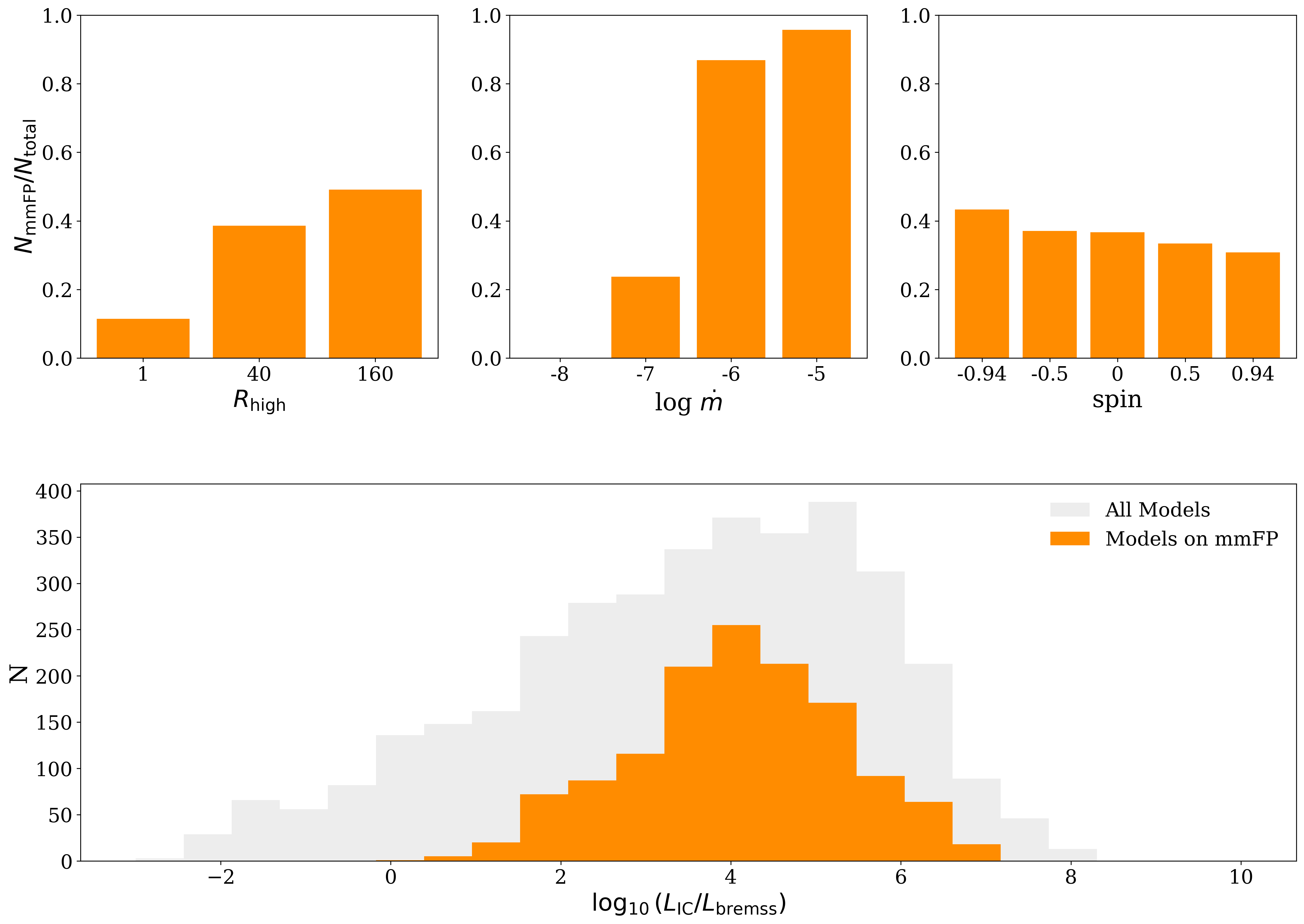}
    \end{tabular}
    \caption{\textit{Top:} Fraction of model snapshots which satisfy mmFP as a function of $R_\mathrm{high}$, Eddington ratio $\dot{m}$ and spin. \textit{Bottom:} Distribution of $\log(L_{\rm IC}/L_{\rm bremss})$ for all models in grey, compared to those consistent with the mmFP in orange.}
    \label{fig:histograms}
\end{figure*}

In \autoref{fig:histograms}, we select the snapshots consistent with the mmFP across $R_{\rm high}$, $\dot m$, spin, and $\log(L_{\rm IC}/L_{\rm bremss})$, where $L_{\rm IC}/L_{\rm bremss}$ is the ratio of Inverse Compton luminosity to bremsstrahlung luminosity only considering 2 to 10 keV emission. As discussed in \autoref{sec:physical_drivers}, models with smaller $\dot{m}$ systematically depart from the relation. All snapshots which satisfy the mmFP have $L_{\rm IC}/L_{\rm bremss}>1$, and only 0.7$\%$ of the total snapshots have $L_{\rm IC}/L_{\rm bremss}<1$. This is consistent with our discussion in \autoref{sec:explain_Xray}. 

We notice a preference for larger $R_{\rm high}$ among mmFP-consistent models, which in the aggregate of models is driven by $\dot{m} = 10^{-7}$.  Larger $\dot{m}$ models typically fall within the scatter regardless of $R_\mathrm{high}$. Because $\nu_{\rm synch}\propto T_e^2 B$ and $\nu_{\rm IC}\propto T_e^4 \dot m^{1/2} M_\bullet^{-1/2}$, increasing $R_{\rm high}$ suppresses the X-rays relative to the millimeter. This helps ``tune'' the intrinsic X-rays into the observed regime without requiring jets or non-thermal electrons, and is consistent with EHT studies that favor large $R_{\rm high}$ to satisfy X-ray upper limits \citep{EHTC+2022e}. 

\section{Discussion}
\label{sec:discussion}

We have computed a grid of SEDs from GRMHD models spanning four parameters, $M_\bullet$, $\dot{m}$, $a_\bullet$, and $R_\mathrm{high}$.  Here, we summarize the effects of each of these parameters in our models.
\begin{itemize}
    \item The black hole mass $M_\bullet$ affects the emitting volume and the scalings of the density and magnetic field. At fixed Eddington ratios, increasing $M_\bullet$ lowers the density ($n \propto \dot{m} M_\bullet^{-1}$), and magnetic field ($B \propto n^{1/2}$), which results in lower synchrotron emissivity per unit volume. However, the total emitting volume grows rapidly ($V \propto M_\bullet^3$) and this more than compensates, resulting in a net scaling of $L_{\nu,mm} \propto \dot{m}^2 M_\bullet$. The X-ray luminosity scales more steeply, as $L_X \propto \dot{m}^{2+\alpha} M_\bullet$ with $\alpha > 0$, reflecting the additional dependence on inverse-Compton scattering and electron temperature $T_e$.
    \item Spin $a_\bullet$ has only a modest influence on our spectra.  The main effect of increasing spin is through the temperature dependence, where higher spins have hotter accretion flows, and progrades are hotter than retrogrades \citep{Dhruv+2025}.  Models with higher spins, and therefore hotter plasma, have higher synchrotron frequencies and more inverse Compton scattering.
    \item Our GRMHD models are in rough qualitative agreement with the semi-analytic ADAF/LLAGN model of \citet{Pesce+2021}, shown in \autoref{fig:total}. Despite its simplicity, the ADAF/LLAGN model qualitatively captures the same trend seen in our GRMHD models; the X-ray emission increases with increasing Eddington ratios. However, it systematically predicts lower X-ray luminosities than our SEDs and is characterized by a single peak in the SED structure instead of two. 
    \item The Eddington ratio $\dot{m}$ increases the bolometric luminosity via $L_\mathrm{bol} \propto \dot{m}^2$, and the IC contribution increases even more steeply via $L_\mathrm{IC} \propto \dot{m}^3$ in the optically thin regime \citep{Narayan+1995}.
    With increasing $\dot{m}$, density and magnetic field strength increases, which in turn pushes the synchrotron and IC peaks to higher frequencies and boosts the overall bolometric luminosity. \citet{Ruffa+2024} reports that $L_{mm}$ is more sensitive to $M_\bullet$ than $\dot{m}$, while we see that models with $\dot{m} < 10^{-6}$ drift to the left of the mmFP. This can be understood as a photosphere effect: for sufficiently large $\dot{m}$, the turnover frequency occurs near or higher than the millimeter.  The total luminosity therefore scales more with the size of the emitting region than with the accretion rate, with the exact scaling dependent on the details of the $n$, $B$, and $T_e$ profiles.
    \item Models with larger $R_{\rm high}$ have lower electron temperatures in the disk by construction. Increasing $R_\mathrm{high}$ therefore decreases the synchrotron and IC power, and moves spectral peaks to lower frequencies. The X-ray component of the spectra is particularly sensitive to this parameter in that $\nu_{IC} \propto T_{e}^{-4}$. \citet{EHTC+2022e} reported that models with larger $R_\mathrm{high}$ could more comfortably pass X-ray upper limits, which can be understood in the context of this strong temperature scaling.
\end{itemize}

We now discuss some physical limitations of this work. All models in this study assume thermal electron distributions. However, non-thermal electron populations, especially in jet or shock associated regions, could produce a high-energy tail of electron energies.  Interestingly, \citet{EHTC+2022e} found that including non-thermal electrons in GRMHD models increased the failure rate due to the X-ray constraint: adding high-energy electrons can easily result in excessive IC, since $\nu_\mathrm{IC} \approx \langle \gamma^2\rangle \nu_\mathrm{seed}$.  As a result, we speculate that including a significant non-thermal electron component may systematically worsen agreement with the mmFP. We emphasize that our use of thermal electron distributions does not rule out non-thermal electron models or jet contributions. The mmFP is defined at millimeter frequencies, where models including only thermal electrons have been successful at reproducing images by the EHT \citep[e.g.,][]{EHTC+2019e,EHTC+2022e}. Our results show that a thermal, disk-dominated model is sufficient to reproduce the observed mmFP, but non-thermal electrons are likely required to explain large-scale radio emission and source sizes at cm wavelengths and longer (e.g., \citet{Issaoun2019}). Non-thermal electrons would preferentially increase emission at high-energies through inverse Compton scattering, increasing $L_X$ relative to $L_{\rm mm}$. This could potentially drive models away from the observed mmFP unless perhaps additional tuning is introduced.

While jets are present in our simulations, their effect is minimized due to our thermal eDF assumption. Classical fundamental plane models \citep{2003MNRAS.343L..59H, 2003ANS...324..453Y, 2005AAS...20710204C} suggest radio and X-ray emission correspond to jet synchrotron and synchrotron self-Compton processes. However, our work illustrates that extended jet emission is not required to reproduce mmFP, at least within the parameter space explored in this study. Adding extended jet emission in the millimeter would systematically shift objects to the right of this relation, if it does not come with the right amount of X-ray emission to cancel this effect.  Size constraints with high enough resolution to distinguish jet from disk would therefore be useful for continuing to probe the physical origins of the mmFP.  For \sgra, the EHT constrains no less than 83\% of the observed emission to come from the compact ring \citep{EHTC+2022b}, while for \m87 the compact flux and jet components are comparable \citep{2019ApJ...875L...4E}.  Extending the EHT array into space with missions such as the Black Hole Explorer (BHEX) will enable horizon-scale images of an additional $\sim$10 objects that lie on the mmFP \citep{2025ApJ...985...41Z}.  Such size constraints will help discriminate this model from others where the millimeter emission originates from much larger radii \citep{Hankla+2025}.

Our work neglects radiative cooling, which is expected to affect the temperature for Eddington ratios as small as $\dot{m}\gtrsim10^{-7}$ \citep{2012MNRAS.426.1928D} and the global dynamics for Eddington ratios $\dot{m}\gtrsim10^{-5.5}$ \citep{Singh+2025}.  Indeed, this latter value roughly matches the transition where our models begin to produce unacceptably high radiative efficiencies due to IC scattering.  Interestingly, radiative GRMHD simulations tend not to produce low enough temperatures (large enough effective values of $R_\mathrm{high}$) to reproduce polarized observations, an area of ongoing development in the field \citep[e.g.,][]{Dihingia+2023,Chael2025}. Cooling reduces electron temperatures and suppresses high-energy emission, particularly at higher $\dot{m}$. Increasing $R_{\rm high}$ is meant to mimic this effect, although this does not capture dynamical feedback on the accretion flow. Setting $R_{\rm low}>1$ would also have achieved a similar effect \citep[e.g.,][]{EHTC+2023,Chan&Chan2025}, although we neglect this parameter to reduce the dimensionality of our library. Incorporating cooling self-consistently would likely lower X-ray luminosities and may improve agreement with observations, at higher accretion rates.

We performed a series of tests on a snapshot with $a_\bullet=0.94$, $R_\mathrm{high}=40$, $M_\bullet=10^{9}M_{\odot}$ and $\dot{m}=10^{-4}$ to assess the impact of several numerical limitations: the magnetization ceiling, the temperature floor, and the finite domain size. First, we applied a ``sigma cut'' by zeroing the density in all cells with plasma $\sigma > 1$. Second, our implementation of \texttt{grmonty} sets a temperature floor of $\Theta_e=10^{-3}$, and we assessed the sensitivity to the electron temperature floor (to any cool, non-relativistic electron populations) by increasing it to 0.3. Third, we tested the impact of the outer torus in the initial conditions by zeroing the density in cells with $r>100 \ r_g$. None of these tests resulted in notable changes to the SED in the bands considered.  

\section{Conclusion}
\label{sec:conclusion}

In this work, we tested a four-dimensional grid of compact, hot accretion flow models with dynamically important magnetic fields against the Millimeter Fundamental Plane (mmFP) of \citet{Ruffa+2024}, an observed correlation between millimeter luminosity, X-ray luminosity and SMBH mass.  Our main results are as follows:

\begin{itemize}
    \item Models with $\dot{m} \gtrsim 10^{-6}$ naturally reproduce the mmFP without any fine-tuning. As $\dot{m}$ decreases below the values exhibited by AGN in the \citet{Ruffa+2024} sample, models drift leftward of the relationship.
    \item X-rays are generated by inverse Compton scattering in all of our time-averaged models. This Compton-produced component exceeds direct bremsstrahlung emission even for models with $\dot{m} = 10^{-8}$.
    \item We explore trends in our models as a function of model parameters.  Spin has only a minor effect on the luminosity in either band, due to higher temperatures achieved with higher spins.  Cooling electrons (with the parameter $R_\mathrm{high}$) has a significant effect on both bands, especially the amount of X-rays produced by IC scattering.
\end{itemize}

We conclude that compact, hot accretion flows with dynamically important magnetic fields naturally reproduce the mmFP.  This work yields two testable predictions:  
\begin{enumerate}
    \item Objects with $\dot{m} < 10^{-7}$ should systematically migrate to the left of the mmFP.  We confirm this behavior in the case of \sgra.
    \item High-resolution size constraints, such as those enabled by EHT and BHEX, should localize the millimeter emitting region to scales of $\lesssim 10-100 \ GM_\bullet/c^2$ for the Eddington ratios of interest.
\end{enumerate}

Our investigation is restricted to $\dot{m}\lesssim 10^{-5}$ due to the lack of radiative cooling in these models.  Future investigation into the impact of weaker magnetic fields, non-thermal electrons, and radiative cooling would help determine the generality of the conclusions reached in this work.

\section{Acknowledgments}
The authors thank Michael D.~Johnson and Daniel Palumbo for informative discussions.  The authors also thank Charles Gammie and Vedant Dhruv for making the GRMHD simulations used in this work publicly available.

This publication is funded in part by the Gordon and Betty Moore Foundation, Grant GBMF-12987.  This publication is supported by the Black Hole Initiative at Harvard University, which is funded by grants from the John Templeton Foundation (Grant \#62286) and the Gordon and Betty Moore Foundation (Grant GBMF-8273)---although the opinions expressed in this work are those of the author(s) and do not necessarily reflect the views of these Foundations. K.M. was supported by École Doctorale 564 Physique en Île-de-France, and G.N.W. was supported by the Princeton Gravity Initiative and the Taplin Fellowship. K.M. was also supported in part through generous support from Mr.\ Michael Tuteur and Amy Tuteur, MD. The authors acknowledge financial support from the National Science Foundation (AST-2307887).

\appendix

\section{Radiative Efficiencies}
\label{sec:radiative_efficiency}

We define the radiative efficiency $\epsilon_{\rm rad}$ as the fraction of rest-mass energy converted into radiation.  For each of our models, we compute their radiative efficiencies by calculating

\begin{equation}
    \epsilon_{\rm rad} = \frac{L_{\rm bol}}{\dot{M}_{\bullet} c^2},
\end{equation}

where $L_{\rm bol} = \int L_\nu d\nu$ is the bolometric luminosity and $\dot{M}_{\bullet} = \dot{m}\dot{M}_\mathrm{Edd}$ is the mass accretion rate.  The results are tabulated in \autoref{tab:rad_eff}.  Our GRMHD simulations are meant to represent radiatively inefficient accretion flows and thus physically consistent models should have $\epsilon_{\rm rad} \ll 1$.  In our paper, we reject all models with $\epsilon_{\rm rad} > 0.1$, which includes all $\dot{m}=10^{-4}$ models.  We find that IC tends to dominate the SED in models with $\epsilon_{\rm rad} > 0.1$; IC would therefore cool the plasma in a more physically complete model. 

\begin{table*}[htb]
\begin{minipage}{0.50\textwidth}
\resizebox{\textwidth}{!}{
\begin{tabular}{|c|c|c|c|c|}
\hline
$a_\bullet$ & $\log M_\bullet$ & $\log \dot{m}$ & $R_\mathrm{high}$ & $\log \epsilon_\mathrm{rad}$ \\
\hline
0.94 & 6 & -5 & 160 & -0.33 \\
0.94 & 6 & -6 & 160 & -1.48 \\
0.94 & 6 & -7 & 160 & -2.49 \\
0.94 & 6 & -8 & 160 & -3.48 \\
\hline
0.94 & 7 & -5 & 160 & -0.34 \\
0.94 & 7 & -6 & 160 & -1.50 \\
0.94 & 7 & -7 & 160 & -2.50 \\
0.94 & 7 & -8 & 160 & -3.49 \\
\hline
0.94 & 8 & -5 & 160 & -0.36 \\
0.94 & 8 & -6 & 160 & -1.52 \\
0.94 & 8 & -7 & 160 & -2.52 \\
0.94 & 8 & -8 & 160 & -3.50 \\
\hline
0.94 & 9 & -5 & 160 & -0.39 \\
0.94 & 9 & -6 & 160 & -1.54 \\
0.94 & 9 & -7 & 160 & -2.54 \\
0.94 & 9 & -8 & 160 & -3.52 \\
\hline
\end{tabular}
}
\end{minipage}
\begin{minipage}{0.50\textwidth}
\resizebox{\textwidth}{!}{
\begin{tabular}{|c|c|c|c|c|}
\hline
$a_\bullet$ & $\log M_\bullet$ & $\log \dot{m}$ & $R_\mathrm{high}$ & $\log \epsilon_\mathrm{rad}$ \\
\hline
0 & 6 & -5 & 160 & -0.83 \\
0 & 6 & -6 & 160 & -1.87 \\
0 & 6 & -7 & 160 & -2.86 \\
0 & 6 & -8 & 160 & -3.84 \\
\hline
0 & 7 & -5 & 160 & -1.84 \\
0 & 7 & -6 & 160 & -2.87 \\
0 & 7 & -7 & 160 & -3.86 \\
0 & 7 & -8 & 160 & -4.85 \\
\hline
0 & 8 & -5 & 160 & -2.84 \\
0 & 8 & -6 & 160 & -3.87 \\
0 & 8 & -7 & 160 & -4.86 \\
0 & 8 & -8 & 160 & -5.85 \\
\hline
0 & 9 & -5 & 160 & -3.83 \\
0 & 9 & -6 & 160 & -4.87 \\
0 & 9 & -7 & 160 & -5.85 \\
0 & 9 & -8 & 160 & -6.84 \\
\hline
\end{tabular}
}
\end{minipage}
\caption{Radiative efficiencies for a subset of our prograde($a_\bullet=+0.94$) and nonspinning($a_\bullet=0$) GRMHD models.}
\label{tab:rad_eff}
\end{table*}

\section{Observational Selection Effects}
\label{sec:selection_effects}
In \autoref{sec:models_vs_data}, we found that models with $\dot{m}<10^{-6}$ systematically drift to the left of the mmFP.  In \autoref{sec:explain_millimeter}, we argued that a strong dependence on $\dot{m}$ is expected, but only when $\dot{m}$ is sufficiently small that the synchrotron turnover frequency $\nu_\mathrm{synch} < 230 \ \mathrm{GHz}$.  Here, we explore the possibility that an observational selection effect explains the difference between our models and the \citet{Ruffa+2024} sample.

\begin{figure*}
    \centering
    \includegraphics[width=
    \textwidth]{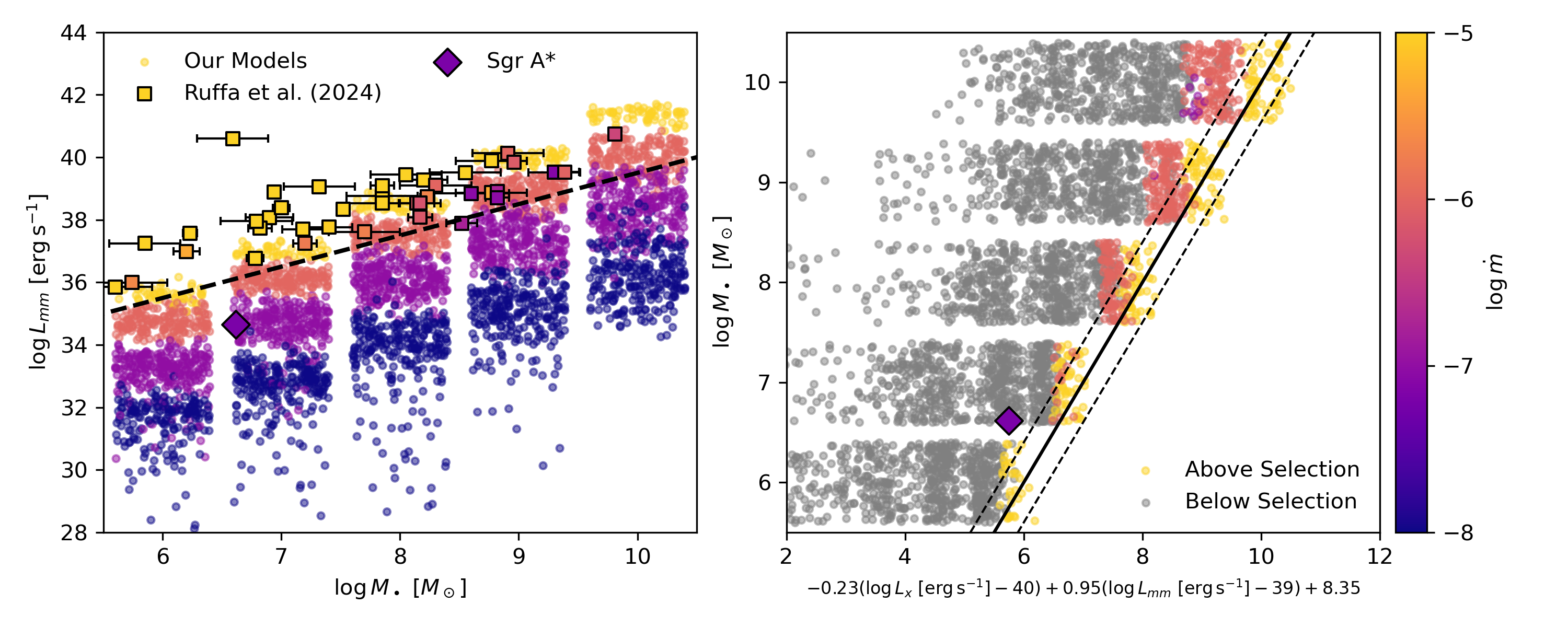}
    \caption{Proposed impact of selection effects.  In the leftmost panel, we compare our GRMHD models to the observational sample from \citet{Ruffa+2024}, color-coded by $\dot{m}$.  For a given $L_{mm}$ and $M_\bullet$, the  \citet{Ruffa+2024} sample generally overlaps with our models in $\dot{m}$, suggesting that our models are producing reasonable millimeter luminosities.  The \citet{Ruffa+2024} sample only includes objects roughly above the dashed black line.  If we restrict our models to those above this line, we find much better agreement with the mmFP in the right panel.  This suggests that objects should drift to the left of the mmFP at lower accretion rates, as the synchrotron turnover frequency moves to lower frequencies.}
    \label{fig:selection_effects}
\end{figure*}

In \autoref{fig:selection_effects}, we first compare our models with the \citet{Ruffa+2024} sample in the $M_\bullet-L_{mm}$ plane.  The \citet{Ruffa+2024} sample spans a broad range in both $M_\bullet$ and $\dot{m}$ (inferred from $L_X$).  Their $\dot{m}$ values exhibit 0th, 25th, 50th, 75th, and 100th percentile values of -7.1, -6.0, -4.9, -3.1, and -1.1 respectively.  All of their objects with $\dot{m} < 10^{-6}$ have $M_\bullet > 10^8 \ M_\odot$.  We color-code both our models and the observational data points by $\dot{m}$, although in the case of the observational data, this is computed assuming $L_X=0.1 L_\mathrm{bol}$ and $\dot{m} = L_\mathrm{bol}/L_\mathrm{Edd}$, which is not generally true in our models.  Note that although our color bar saturates at $\dot{m}=10^{-5}$, about half of the \citet{Ruffa+2024} objects lie above that value.  Comparing the models with the observed data, our models clearly extend to far smaller values of $L_{mm}$ than the observed sample.  At fixed $L_{mm}$ and $M_\bullet$, the observations and models roughly agree on $\dot{m}$, implying that our models are producing reasonable millimeter luminosities, but the observed sample only overlaps with the upper envelope of our models.

The \citet{Ruffa+2024} objects fall roughly above the dashed line defined by $\log L_{mm} \ [\mathrm{erg}\,\mathrm{s}^{-1}] = 35.5 + (\log M_\bullet [M_\odot]-6)$.  As shown in the second panel, if we select for the models that lie above this line, they agree very well with the mmFP.  That is, if we select for models with comparable millimeter luminosities to the observed sample, then they naturally produce an appropriate amount of X-ray emission via IC scattering.  This suggests that an observational selection effect leads to the disagreement of our low $\dot{m}$ models with the mmFP.  Indeed, when we include \sgra on these plots, with $L_x = 2.4 \times 10^{33} \ \mathrm{erg}\,\mathrm{s}^{-1}$ \citep{Baganoff+2003}, $M_\bullet=4.14\times 10^6 \ M_\odot$ \citep{Do+2019}, $L_{mm}=4.6 \times 10^{34} \ \mathrm{erg}\,\mathrm{s}^{-1}$ \citep{EHTC+2022c}, and  a ``best bet'' estimate of $\dot{m} \approx 10^{-7.2}$ \citep{EHTC+2022e,EHTC+2024c}, we find that it agrees well with our models but has much lower $L_{mm}$ than the \citet{Ruffa+2024} sample and falls to the left of this relationship as expected.  A larger sample of extremely low Eddington ratio objects could therefore place interesting constraints on the physical origins of the mmFP.

\hfill
\bibliography{ms}

\end{document}